\documentclass[%
 reprint,
superscriptaddress,
 amsmath,amssymb,
 aps,
]{revtex4-1}

\usepackage{graphicx}
\usepackage{dcolumn}
\usepackage{bm}
\usepackage{siunitx}

\newcommand{\CUO}{Cu$_x$O$_y$\,}

\newcommand{\mat}[1]{\boldsymbol{\mathrm{#1}}} 

\begin{document}

\title{Ultrathin 2\,nm gold as ideal impedance-matched absorber for infrared light}

\author{Niklas Luhmann}
\affiliation{Institute of Sensor and Actuator Systems, TU Wien, 1040 Vienna, Austria.}
\author{Dennis H\o j}
\affiliation{Department of Physics, Technical University of Denmark, 2800 Kongens Lyngby, Denmark}
\author{Markus Piller}
\affiliation{Institute of Sensor and Actuator Systems, TU Wien, 1040 Vienna, Austria.}
\author{Hendrik K\"ahler}
\affiliation{Institute of Sensor and Actuator Systems, TU Wien, 1040 Vienna, Austria.}
\author{Miao-Hsuan Chien}
\affiliation{Institute of Sensor and Actuator Systems, TU Wien, 1040 Vienna, Austria.}
\author{Robert G. West}
\affiliation{Institute of Sensor and Actuator Systems, TU Wien, 1040 Vienna, Austria.}
\author{Ulrik Lund Andersen}
\affiliation{Department of Physics, Technical University of Denmark, 2800 Kongens Lyngby, Denmark}
\author{Silvan Schmid}%
 \email{silvan.schmid@tuwien.ac.at}
\affiliation{Institute of Sensor and Actuator Systems, TU Wien, 1040 Vienna, Austria.}

\date{\today}

\begin{abstract}
Thermal detectors are a cornerstone of infrared (IR) and terahertz (THz) technology due to their broad spectral range. These detectors call for suitable broad spectral absorbers with minimal thermal mass. Often this is realized by plasmonic absorbers, which ensure a high absorptivity but only for a narrow spectral band \cite{Liu2015, Ma2013, Hui2016}. Alternativly, a common approach is based on impedance-matching the sheet resistance of a thin metallic film to half the free-space impedance. Thereby, it is possible to achieve a wavelength-independent absorptivity of up to \SI{50}{\percent}, depending on the dielectric properties of the underlying substrate \cite{Lang1992, Bauer1993, Mattiucci2013}. However, existing absorber films typically require a thickness of the order of tens of nanometers, such as titanium nitride (\SI{14}{\nano\meter}) \cite{Laurent2018}, which can significantly deteriorate the response of a thermal transducers. Here, we present the application of ultrathin gold (\SI{2}{\nano\meter}) on top of a \SI{1.2}{\nano\meter} copper oxide seed layer as an effective IR absorber. An almost wavelength-independent and long-time stable absorptivity of \SI{\sim 47\pm 3}{\percent}, ranging from \SIrange{2}{20}{\micro\meter}, could be obtained and is further discussed. The presented gold thin-film represents an almost ideal impedance-matched IR absorber that allows a significant improvement of state-of-the-art thermal detector technology.

\end{abstract}


\maketitle

\section{Introduction}

Thermal IR/THz detectors have remained the only technology covering the entire spectral range from the visible all the way to the far-IR (THz) regime, facilitating applications ranging from spectrochemical analysis to security and astronomy \cite{Rogalski_2019,Minoglou_2019, Liu_2019,Gerlach_2019,Talghader_2018,G_n_reux_2018}. Thermal detectors exploit the conversion of the absorbed photothermal power to either a change in electrical resistance or electric potential, as in bolometers or pyroelectrics and thermocouples, respectively \cite{Rogalski_2019}. More recently, micro- and nanoelectromechanical systems (MEMS/NEMS) have demonstrated exceptional potential as IR and THz detectors due to strong photothermally-induced detuning of their mechanical resonance frequency \cite{Piller, Blaikie_2019, Laurent2018, Zhang2013, Hui2016, Zhang2016, Zhang2019, Gokhale2014}. 

Despite such novel variations in detection paradigms, the development of efficient, broadband photothermal transducers remains a crucial task in high-performance IR/THz detection. An \textit{ideal absorber}, in this case, should provide long-term stability and a broad spectral response while having an negligible thermal mass \cite{Parsons1988}, which is given by $C_{th} = n\, c_m$.
Here $n$ is the total number of atoms and $c_m$ the molar heat capacity of the material. Therefore, since $c_m$ is more or less constant for metals, minimizing the number of absorber atoms, that is the film thickness for a given absorber area, is the only way to minimize $C_{th}$. This has been the focus for advancement in IR/THz detection for decades and has led to a reduction in effective thickness of the absorber to the order of 10~nm in the present day \cite{Rogalski_2019}.

At this scale, contemporary solutions are numerous: from novel antenna structures \cite{Ma2013, Hui2016,Yu_2016,Gou_2018} to metamaterials \cite{Yin2015,Takagawa_2015,Jin2012}, which promise a high absorptivity but are limited by their resonance bandwidth. A progressive solution to improve the spectral range of detection has been to use a stack of plasmonic structures with differing lateral size; nonetheless, the bandwidth of these sensors remains limited \cite{Liu2015}. A most recent thrust toward the ultimate limit of uncooled detection has been to employ the exceptional sensitivity of graphene \cite{Rogalski_2019b}, even making the detector itself the primary absorber, for example, as an uncooled NEMS resonator \cite{Blaikie_2019}. In respect to ultra thin layers for absorption, two-dimensional (2D) materials as graphene are currently of great interest in research due to its fascinating properties which make new detection concepts possible \cite{Wang2019, Rogalski_2019}.
Graphene can be utilized in thermal or photon detection, although most research and development has been focused on its unique photovoltaic properties for photon detection, the spectral absorptivity is limited \cite{Rogalski_2019b, Rogalski_2019, Gul2019}.
Furthermore, due to graphene's low absorptivity in the near- to mid-IR of only \SI{2.3}{\percent} \cite{Yan2011}, modifications using plasmonic metastructures are still required \cite{Shimatani_2019}, and the bandwidth limitation remains an issue in modified graphene absorbers \cite{Gul_2019}.  

In the modern age of nano- and atomic-scale detectors, we may need to return to a classic, old-fashioned approach: to engineer the sheet resistance of a thin metal such that it matches half the free space impedance of \SI{\sim 377}{\ohm} \cite{Hagen_1902,Hagen_1903,Woltersdorff1934}. Based on the theory introduced by Hadley \cite{HADLEY1947} and C. Hilsum \cite{Hilsum1955}, a \textit{wavelength-independent} absorptivity of up to \SI{50}{\percent} can be achieved, assuming the optical constants $n,k$ are approximately equal --- which, for metals such as gold, is only valid in the far-IR.

Many approaches using thin layers of e.g. chromium oxide, silver or platinum \cite{Kruse2001, Ajakaiye2007, Lee2001, Bauer1993, Formica_2013, Lang1992, Mahan1983}, unseeded metastructures \cite{Kim_2016}, and alloys, such as titanium nitride \cite{Zhang2013,Laurent2018,K_V__2019}, have been successfully tested for this purpose. However, some alloys and metals are prone to oxidization, which changes the absorptivity over time \cite{Drobny_1979,Formica_2013}. Regarding the thickness needed to match the desired impedance, alloys such as TiN, with an optimum of \SI{14}{\nano\meter}, are often in the same dimension as the detecting element itself \cite{Laurent2018,Piller,Zhang2013}. Other metals, such as gold (Au), demonstrate strong percolation effects, setting a lower limit for thin-film thickness at the insulator-to-metal transition, which normally makes them too conductive to match the necessary sheet resistance \cite{Hovel2010}. 


This issue has been addressed by using for instance copper as a seed layer to generate atomically smooth thin films of silver \cite{Formica_2013}. Recent studies on tunable plasmons presented a similar methodology to fabricate ultrathin metallic films of Au below its typical percolation of \SI{\sim 7}{\nano\meter} \cite{Maniyara2019}. With help of a surfactant layer of copper oxide (\CUO) of \SI{\sim 1}{\nano\meter}, it is possible to reduce the percolation threshold of gold to less than \SI{2}{\nano\meter}. Here, we demonstrate the application of this technique to fabricate an ultrathin Au layer as an efficient, broad spectral impedance-matched IR absorber. With respect to the deviated optical properties of ultra-thin gold compared to bulk values, we found an almost wavelength-independent absorptivity down to below \SI{2}{\micro\meter}.

\section{Methods}

\begin{figure}[t]
    \centering
    \includegraphics[width=0.45\textwidth]{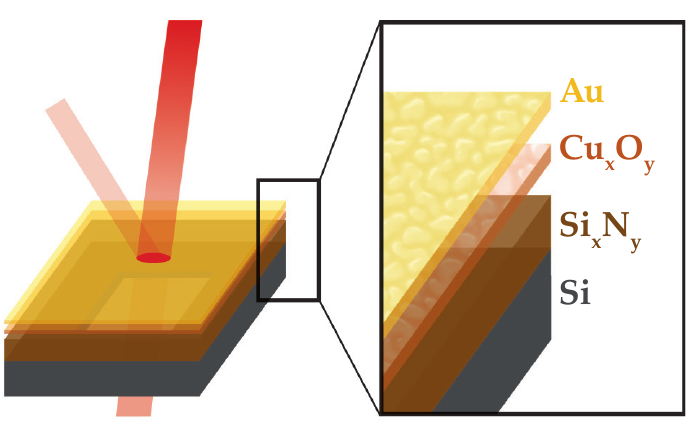}
    \caption{Illustration of the fabricated layers and probing direction. The samples are based on \SI{50}{\nano\meter} Si$_x$N$_y$ membranes comprising a \SI{1.2(2)}{\nano\meter} sputtered and naturally oxidized \CUO layer. For characterization of optimal thickness, varying Au layers were deposited on top using physical vapour deposition.}
    \label{fig:Schematic}
\end{figure}

Figure \ref{fig:Schematic} shows a schematic illustration of the fabricated samples used for this study. All experiments were conducted on \SI{50}{\nano\meter} thin silicon nitride (Si$_x$N$_y$) deposited by low-pressure chemical vapour deposition (LPCVD), acting as a support layer with minimal absorptivity. The \SI{2.5x2.5}{\milli\meter} squared membranes were structured by standard UV lithography and backside released in potassium hydroxide (KOH). For the ultrathin alloy fabrication, a Cu layer was deposited by sputter deposition using a 'Von Ardenne LS 730 S'. The deposition rate was set to \SI{1.5}{\angstrom\per\second}, extrapolated from several test depositions, resulting in a seed layer thickness of \SI{1.2(2)}{\nano\meter}. To ensure a smooth and clean surface, the Si$_x$N$_y$ membranes were plasma cleaned using Argon-based reactive ion etching in the same vacuum chamber, immediately before Cu deposition. Following Maniyara et al.\cite{Maniyara2019}, the samples were then stored in air for one day to undergo oxidation. In a final step, gold was evaporated from a tungsten boat with a comparably low rate of \SI{0.3}{\angstrom\per\second} at \SI{3e-8}{\milli\bar}. The deposition rate and extracted nominal thickness was monitored by a quartz resonating sensor. 

All optical spectra were recorded via Fourier Transformed IR spectroscopy (FTIR) conducted with a 'Bruker Tensor 27'. In order to minimize systematic deviations, transmittance and reflectivity measurements were performed within one measurement using a specified 'A 510/Q-T' set-up and an aperture of \SI{2}{\milli\meter}.

The resistivity and sheet resistance were obtained by a homemade four-point probe setup made of a cylindrical probe-head provided by 'Jandel', a 'Keithley 6221' current source, and a 'Keithley 2182A' nanovoltmeter. The probe-head was attached to a load bending beam 'Burster 8511-5050' to monitor the contact force applied on the surface during measurements. All samples were measured in a current range of $10^{-7}$ to $10^{-3}$\,A and maximum contact force of \SI{2.2}{\newton}.   

\section{Theory}

As mentioned, L. Hadley \cite{HADLEY1947} and C. Hilsum \cite{Hilsum1955} showed thin metal sheets can obtain \SI{50}{\percent} absorptivity in infrared by matching its sheet resistance with half the free space impedance $\sqrt{\mu_0/4\varepsilon_0} \approx 188 \ \Omega$, where $\mu_0$ and $\varepsilon_0$ are the permeability and permittivity of free space, respectively. Here, an important assumption made is that the refractive index $n$ and extinction coefficient $\kappa$ of the metal are to be equal. From our knowledge the validity of this criteria has yet not been discussed, especially in regard to the limiting wavelength where it can be applied. 

To get a better understanding of when this assumption is valid, the Drude model is used and rewritten in terms of plasma frequency $\omega_p$ and electrical resistivity $\rho$:
\begin{equation}
\hat{\varepsilon} = \varepsilon_1 + \mathrm{i}\varepsilon_2 = 1 - \frac{1}{\omega^2/\omega_p^2 + \mathrm{i}\varepsilon_0\,\rho\,\omega}
\label{eq:Drude-Model}
\ ,
\end{equation}
where $\omega$ is the angular frequency of the optical field. The refractive index and extinction coefficient is then given by $n = \sqrt{|\hat{\varepsilon}| + \varepsilon_1}$ and $\kappa = \sqrt{|\hat{\varepsilon}| - \varepsilon_1}$, respectively. It can clearly be seen, in order for Hilsum/Hadley's assumption to be valid, the imaginary part of the complex relative permittivity $\varepsilon_2$ must dominate. This is true when 
\begin{equation}
    \omega \ll \varepsilon_0\,\rho\,\omega_p^2
    \ .
    \label{eq:HilsimCrit}
\end{equation}
In this regime, the Drude model can then be simplified to
\begin{equation}
    \hat{\varepsilon} \approx \frac{1}{\varepsilon_0^2\,\rho^2\,\omega_p^2} + \mathrm{i}\frac{1}{\varepsilon_0\rho\,\omega}
    \label{eq:simpl-Drude-Model}
    \ ,
\end{equation}
where it can be seen that the real part is limited to some finite value; whereas, the imaginary part is increasing for longer wavelengths. For gold, assuming bulk values $\rho = \SI{2.2e-8}{\ohm\!\cdot\!\meter}$ and $\omega_p = 2\pi \cdot \SI{2.1}{\peta\hertz}$, this limit is approximately at a wavelength of \SI{56}{\micro\meter}. Below this limit it should not be possible to achieve high absorptivity, unless the material parameters change, which is indeed the case for ultrathin metal films \cite{Hovel2010}. Especially regarding ultrathin layers, the resistivity can be many factors of magnitude higher than compared to bulk \cite{Schmiedl2008, Wu2004}. To a certain extent, this can be described by the so-called scattering hypothesis; whereby, the material's resistivity is defined as a sum of scattering contributions \cite{Wissmann2007}
\begin{equation}
    \rho=\rho_0+\rho_{GB} + \rho_{SS} + \rho_{SR}
    \ ,
    \label{eq:resistivity}
\end{equation}

where $\rho_0$ is the bulk resistivity, $\rho_{GB}\propto D^{-1}$ is the grain boundary contribution, $\rho_{SS}\propto d^{-1}$ is the surface scattering contribution, and $\rho_{SR} \propto d^{-3}$ is the roughness contribution. Here, $d$ is the metal thickness and $D$ the mean grain size of the metal film. For thin films, one can approximate $D$ to be equal to $d$ \cite{Bandyopadhyay1979}. However, with increasing thickness, a limiting grain size of $D_\infty$ is reached. The grain-boundary contribution can therefore be extended to $\rho_{GB} \propto 1/D_\infty + 1/(Cd)$, where $D_\infty$ is often found to be limited up to a thickness of \SI{\sim\,20}{\nano\meter},  the range of the materials electron mean free path. The dimensionless factor $C$, typically ranges from 0.5 to 1 \cite{Sambles1982}. Below percolation, ohmic bridges and tunneling effects govern the resistivity, which can be described by e.g. Monte-Carlo simulations or a filamentary vibron quantum percolation model \cite{Finzel1985, Phillips2001}, but are beyond the scope of this study. Collectively, all these resistivity contributions lead to broaden the absorptivity bandwidth of thin films down to shorter wavelengths as described by (\ref{eq:HilsimCrit}).

In order to model the optical behaviour, a general matrix method has been implemented \cite{Centurioni2005} to predict the transmittance and reflectivity (see appendix \ref{app:OpticalModel}). The incident light is assumed to be non-polarized and normal to the metal surface. The material layers are assumed to be coherent, e.g. isotropic and homogeneous. Assuming all copper has been oxidized, hence not conductive, it will have negligible contribution and is, therefore, excluded from the model. Optical data for Si$_x$N$_y$ and Au are extracted experimentally which will be described in the following section.\\

\section{Results and Discussion}

\begin{figure}[t]
    \centering
    \includegraphics[width=0.45\textwidth]{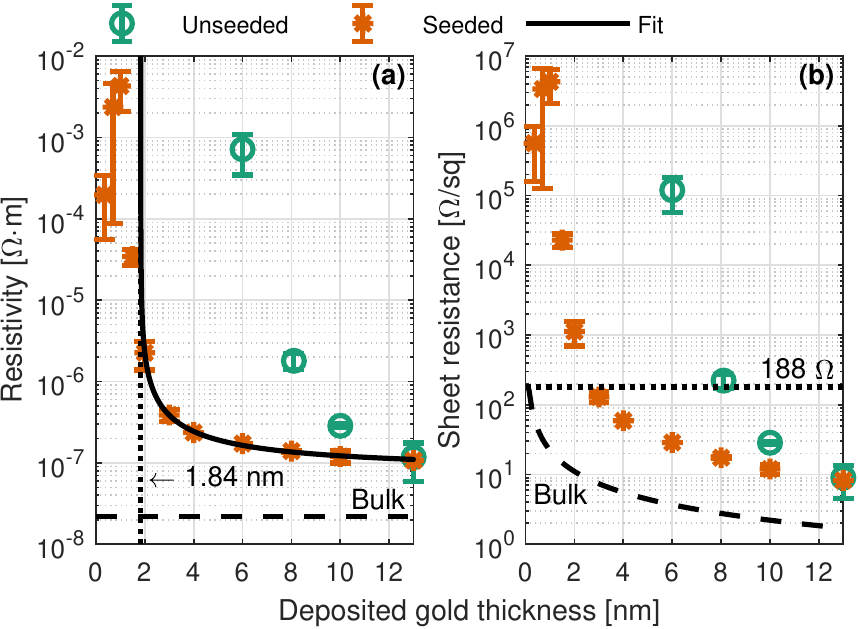}
    \caption{(a) Measured resistivity of seeded and unseeded Au layers as a function of deposited thickness. Due to the \CUO seed layer a metallic behaviour of Au can be obtained down to \SI{2}{\nano\meter}. The resistivity can be well described by the scattering hypothesis (\ref{eq:resistivity}) and is strongly governed by grain-boundary and surface scattering $\propto d^{-1}$. In order to fit the data, a \SI{1.8}{\nano\meter} offset is included to compensate the percolation threshold and uncertainty of effective thickness.  (b) Sheet resistance of the same samples. The optimal thickness for an ideal impedance-matched is expected at \SI{2.5}{\nano\meter}.  Bulk gold data taken from \cite{Haynes2010}.}
    \label{fig:R_vs_h}
\end{figure}

Figure \ref{fig:R_vs_h} shows the measured resistivity and corresponding sheet resistance of seeded vs. unseeded gold as a function of the deposited layer thickness. Consistent with previous studies made on ultrathin copper and gold films \cite{Schmiedl2008, Hovel2010, Henriquez2019}, the resistivity can be partly fitted by the scattering hypothesis (\ref{eq:resistivity}) including a variable offset to compensate the percolation threshold and uncertainty of the effective film thickness. The data can be well described by the model down to \SI{2}{nm}, including a positive offset of \SI{1.8}{\nano\meter} and is governed by the grain boundary and surface scattering term $\propto d^{-1}$. The offset can be mainly related to the percolation threshold. Here, the \SI{1.5}{\nano\meter} layer showed a resistivity of \SI{3580}{\micro\ohm\!\cdot\!\centi\meter} while the \SI{2}{\nano\meter} already dropped to \SI{253}{\micro\ohm\centi\meter}. This falls exactly into the insulator-to-metal transition region of \SIrange{600}{200}{\micro\ohm\!\cdot\!\centi\meter}, defined by \textit{Ioffe} and \textit{Regel} \cite{Ioffe1960}. Thus, percolation must occur in between those samples. Considering that the used thickness values are taken from the quartz sensor, an additional uncertainty of $\pm\SI{0.3}{\nano\meter}$ can be expected.

Compared to the bulk resistivity of gold with \SI{2.2e-8}{\ohm\!\cdot\!\meter}, the \SI{13}{\nano\meter} converge to a three times higher value. As mentioned in the theory section, the grain-boundary scattering term includes a limiting grain size factor $1/D_\infty$, which significantly lifts the resistivity up to a thickness in the range of the electron mean free path \SI{\sim\,20}{\nano\meter}. Thus, with respect to previous studies \cite{Sambles1982}, such an increased value can be expected.

Regarding the measured sheet resistance, the \textit{optimal} thickness for matching with half the free space impedance (\SI{188}{\ohm}) is expected around \SI{2.5}{\nano\meter} Au. 

\begin{figure}[t]
    \centering
    \includegraphics[width=0.45\textwidth]{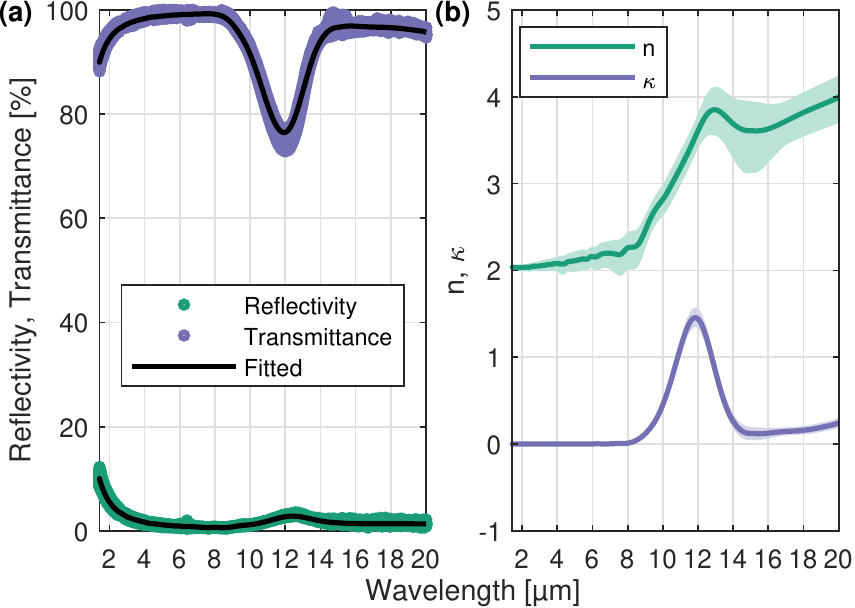}
    \caption{(a) Average of ten individual FTIR measurements on \SI{50}{\nano\meter} silicon nitride membranes. (b) Fitted optical constants using the matrix method \cite{Centurioni2005} by grouping measured data into fixed wavelength points, from which the optical constants were estimated individually (see text). The shaded region indicates the uncertainty estimated as a \SI{50}{\percent} increase in sum of squared residuals.}
    \label{fig:SiN_fit}
\end{figure}

To investigate whether there is a potential contribution of the \CUO seed layer, additional experiments with varying Cu thicknesses revealed, that it could be also used as an impedance-matched absorber; but only if measured directly after deposition. However, due to oxidization over time \cite{Drobny_1979}, all reference samples with Cu - \CUO showed a significant increase of resistivity back to the insulating state, and consequential loss of absorptivity. Regarding the presented data, we found no significant influence on the electrical or optical properties by the seed layer used.\\

In a next step the optical properties of ultrathin Au layers are discussed. Therefore, the transmittance and reflectivity for different (seeded) Au thicknesses were measured and fitted by the described Drude Model, based on (\ref{eq:Drude-Model}). To obtain a proper fit, optical data for the LPCVD Si$_x$N$_y$ was needed. This was extracted from multiple FTIR measurements on bare LPCVD Si$_x$N$_y$ membranes shown in figure~\ref{fig:SiN_fit}(a). Using the above-mentioned optical model adapted to a single Si$_x$N$_y$ layer to predict the transmittance and reflectivity, a nonlinear fit was done at each wavelength separately, in order to estimate the optical constants. To reduce the uncertainty of the fit, the measured data were grouped into spectral blocks with \SI{50}{\percent} overlap. This increases the amount of data per spectral point at the cost of spectral resolution. The result is shown in figure \ref{fig:SiN_fit}(b) with 618 spectral points. Note that the fit is not Kramers-Kronig constrained \cite{DeL.Kronig1926}, which was not possible due to the limited spectral range of the measurements. However, for the purpose of this paper the data is deemed acceptable.

\begin{figure}[t]
    \centering
    \includegraphics[width=0.45\textwidth]{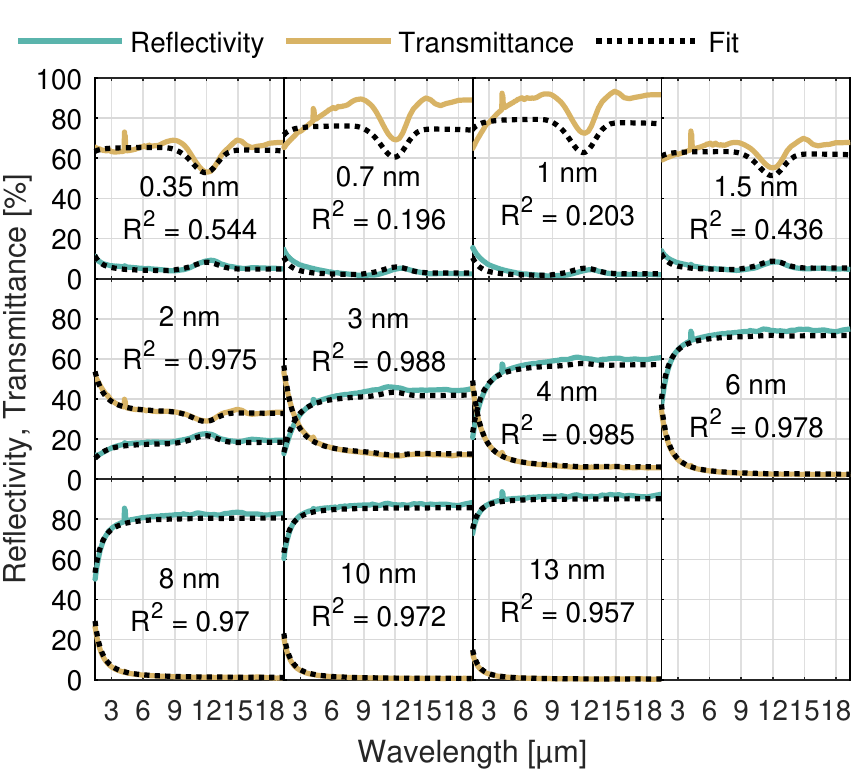}
    \caption{Measured transmittance and reflectivity of (seeded) Au layers, obtained by FTIR spectroscopy. All spectra are fitted by the given Drude Model (\ref{eq:Drude-Model}), including the measured resistivity. $R^2$ is the coefficient of determination for each fit. As for thicker layers, all spectra are in good agreement with the optical properties of a metallic film, the layers below \SI{2}{nm} show a divergence, due to the insulator-to-metal transition. For the 'metallic' layers, the obtained plasma frequency remains constant with  $w_p \sim 2\pi\cdot\SI{3.0\pm 0.7 e15}{\peta\hertz}$.}
    \label{fig:FitPlots}
\end{figure}

Figure~\ref{fig:FitPlots} shows the measured transmittance and reflectivity of all seeded gold layers. The data is fitted by using the optical model based on the matrix method \cite{Centurioni2005}, in combination with the Drude model from (\ref{eq:Drude-Model}), and including the previous extracted optical properties for Si$_x$N$_y$. As can be seen from the $R^2$ coefficient of determination, the thicker layers show a metallic behaviour and can be well fitted down to \SI{2}{\nano\meter}, consistent with the percolation threshold obtained before. For those thicknesses, the extracted plasma frequency with $w_p \approx 2\pi\cdot\SI{3\pm 0.7 e15}{\peta\hertz}$ is slightly increased compared to bulk but remains constant within its uncertainty. Due to the insulating behaviour, the model can not effectively be applied below percolation (\SI{2}{\nano\meter}), which can be clearly seen by the dropping $R^2$ coefficient. In this region, an increase of the transmittance from the \SI{0.35}{nm} to the \SI{0.7}{nm} and \SI{1}{nm} sample can be observed. Previous studies made on Au layers around the percolation threshold have confirmed this 'antireflection' phenomena \cite{Gompf2007}, whose origin lies in the divergence of the dielectric constant $\varepsilon_1$ in that region \cite{Hovel2010}. One could, therefore, suggest the use of such ultrathin Au below percolation as potential antireflection coating, which, however, exceeds the scope of this study.   

\begin{figure}[t]
    \centering
    \includegraphics{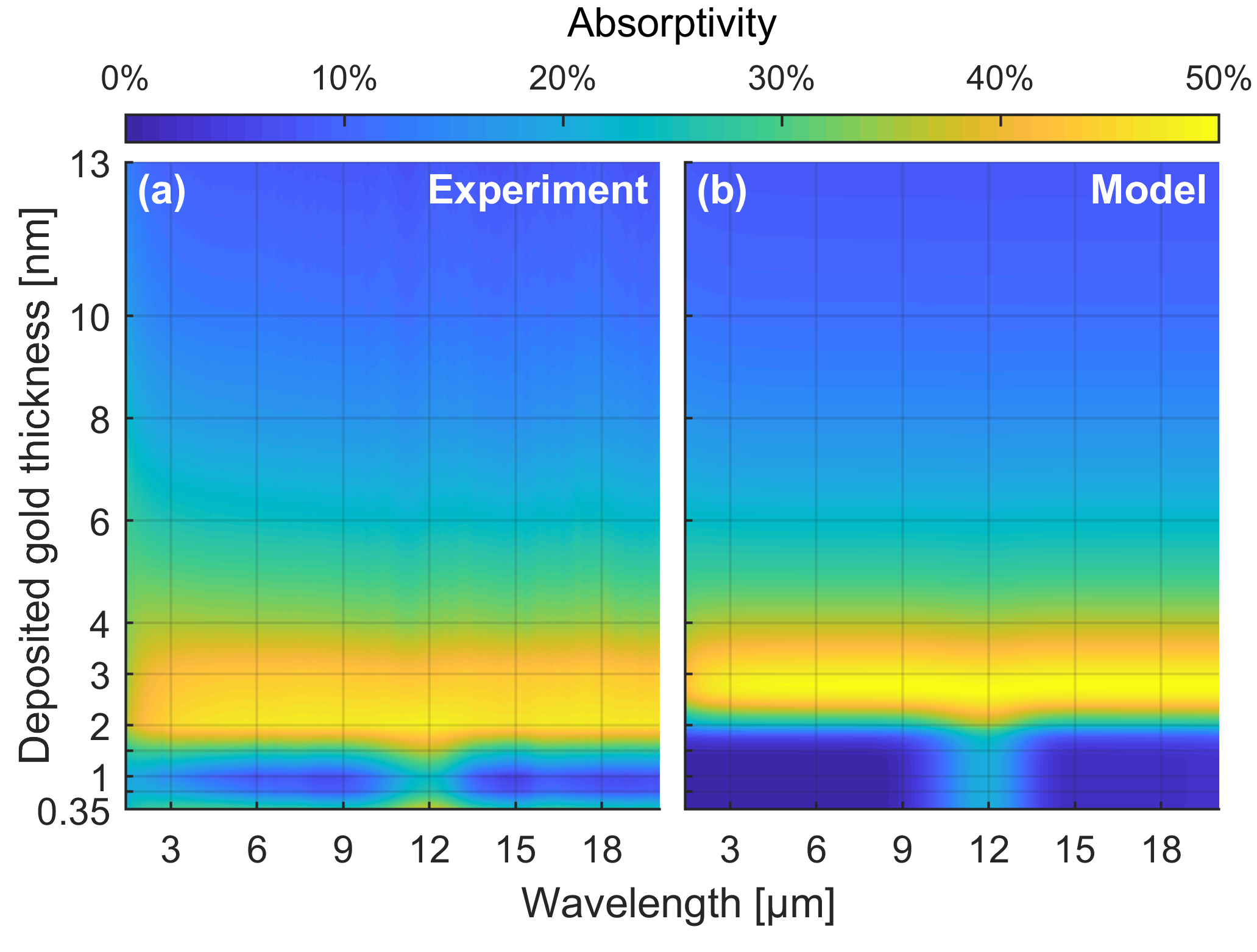}
    \caption{Absorptivity vs deposited gold thickness and wavelength. (a) Experimental obtained absorptivity from FTIR measurements. The data has been linearly interpolated for the plot. Each horizontal grid line corresponds to a sample. (b) Simulated absorptivity using the fitted Drude parameters and extracted optical properties of Si$_x$N$_y$.}
    \label{fig:exp_vs_theory}
\end{figure}

Figure~\ref{fig:exp_vs_theory}(a) shows the measured, linearly interpolated absorptivity (1 $-$ reflectivity $-$ transmittance), plotted over the wavelength and deposited Au thickness. All measured layers were evaporated on \SI{50}{\nano\meter} Si$_x$N$_y$ and a \SI{1.2(2)}{\nano\meter} \CUO seed layer. One can clearly determine a maximum absorptivity of \SIrange{40}{50}{\percent} between the \SI{2}{\nano\meter} and \SI{4}{\nano\meter} Au layer, with a slight decrease towards smaller wavelengths. The absorption peak with a maximum at \SI{12}{\micro\meter} is typical for Si$_x$N$_y$ but becomes dominated by the absorber above \SI{2}{\nano\meter}.   

Since the bulk optical constants are apparently not valid for thin layers, the imaginary part of permittivity was again calculated from the measured resistivity assuming gold as the sole metallic layer. Together with the extracted plasma frequency and modelled optical properties for Si$_x$N$_y$, one can estimate the absorptivity by using the optical model (see Appendix~\ref{app:OpticalModel}). The result is presented in figure~\ref{fig:exp_vs_theory}(b).  A direct comparison to the data shows, in general, a good agreement with a minimal offset of $\sim$ \SI{0.35}{\nano\meter} towards the optimal thickness, which lies again in the uncertainty of quartz sensor. With respect to the derived criteria (\ref{eq:HilsimCrit}), the increased resistivity and constant plasma frequency lead to an almost wavelength independent high absorptivity, ranging from \SI{2}{\micro\meter} up to the detection limit of \SI{20}{\micro\meter}. According to theory, it is expected that the given absorptivity remains constant also in the far-IR.

\begin{figure}[t]
    \centering
    \includegraphics[width=0.45\textwidth]{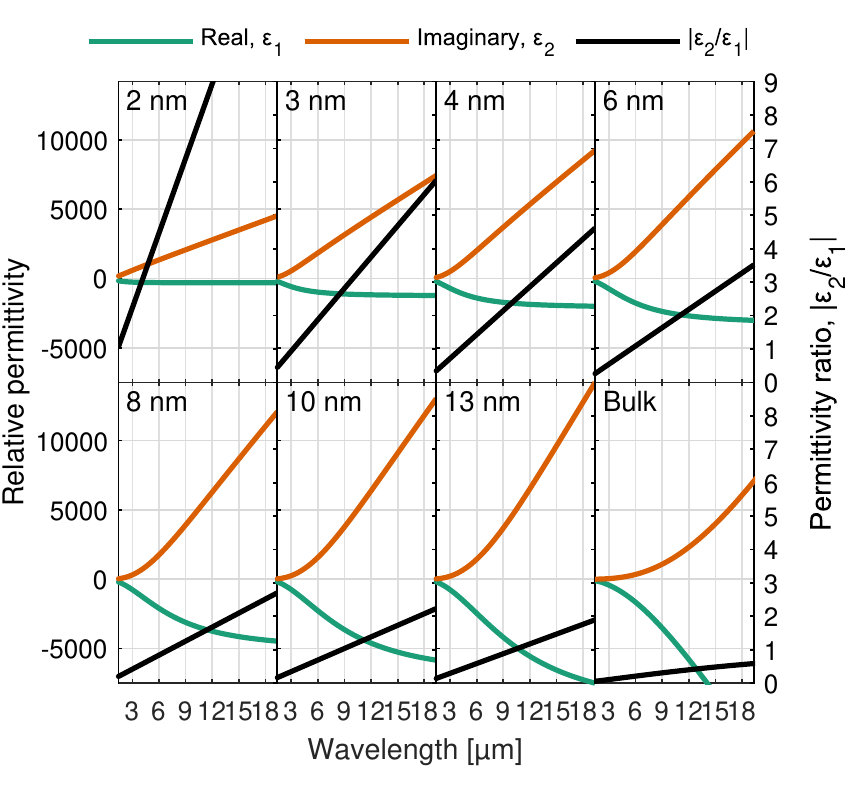}
    \caption{Extracted relative permittivity for increasing gold layer thickness, based on the fitted Drude model (\ref{eq:simpl-Drude-Model}). The solid black line represents a ratio of the imaginary to real part of the permittivity. For thinner layers, where impedance-matched absorption occurs, the imaginary part strongly dominates with increasing wavelength, in good agreement with the derived criteria (\ref{eq:HilsimCrit}). The bulk data is for evaporated gold taken from \cite{Olmon2012}. }
    \label{fig:perm_n_plots}
\end{figure}

In a final step, the dielectric functions of each metallic layer is extracted via (\ref{eq:Drude-Model}) and plotted in figure~\ref{fig:perm_n_plots}. Overall, one can observe a clear dominance of the imaginary part $\varepsilon_2$ compared to the real part $\varepsilon_1$ of the complex permittivity. Moreover, for the \SI{2}{\nano\meter} we obtain an almost negligibly real part and increasing imaginary part towards smaller frequencies and longer wavelength, respectively, in good agreement with the derived criteria (\ref{eq:HilsimCrit}). As can be clearly seen by the ratio $\varepsilon_2/\varepsilon_1$ (black line), this trend can be observed down to short wavelengths, where the real part of the permittivity becomes significant in relation to the imaginary part. Consequentially, this determines the lower limit of the fabricated impedance-matched absorber to \SI{2}{\micro\meter}. Regarding the criteria itself, it should be possible to broaden the absorptivity by using a material with a larger plasma frequency and higher resistivity such as e.g. aluminum, or by other engineered materials such as doped semiconductors, which needs to be further investigated. Knowing the optimal thicknesses and bulk densities of other metals that are typically used for impedance-matched absorption like e.g. platinum or titanium nitride, one can roughly estimate the gain in efficiency due to the reduced thermal mass. Based on previous studies, the \textit{optimal} thickness for impedance-matched absorption were found to be around \SIrange{5}{7}{\nano\meter} for Pt \cite{Mahan1983, Kruse2001, Ajakaiye2007} and \SI{14}{\nano\meter} for TiN \cite{Laurent2018}. Considering a 2.5 mm x 2.5 mm squared membrane, bulk densities and known optimal thicknesses, the presented ultrathin \SI{2}{\nano\meter} Au absorber has an up to 4$\times$ and 9$\times$ times smaller thermal mass compared to impedance-matched absorbers made of Pt or TiN, respectively. With focus on the stability of the fabricated absorber, a re-measurement of the \SI{2}{nm} sample after five month being exposed to air showed no significant change in the nominal absorptivity. Thus, the \SI{2}{\nano\meter} Au seems to be sufficiently stable to prevent the surface from any chemical change.       

\section{Conclusion}

We have demonstrated the application of ultrathin \SI{2}{\nano\meter} Au as a highly efficient, broad spectral, impedance-matched absorber. In good agreement with theoretical assumptions, it was possible to gain  \SI{47(3)}{\percent} absorptivity over the entire near- and mid-IR range from \SIrange{2}{20}{\micro\meter}. According to the theory, this wavelength-independent absorptivity is expected to be also valid in the far-IR. Electrical and optical analysis of the deposited Au layers demonstrated the significantly increased resistivity and impact on the optical properties of such ultrathin metal films, which broaden the lower limit of impedance-matched absorption to \SI{2}{\micro\meter}. In this context, it was possible to approximately match the optimal sheet resistance of about \SI{188}{\ohm}. The extracted dielectric functions verified, that for those samples the imaginary part of the permittivity is strongly enhanced compared to the real part, which in this region can be almost neglected. In the course of this study the optical constants of LPCVD Si$_x$N$_y$ in the range of \SIrange{2}{20}{\micro\meter} were obtained for the first time and further used to estimate the absorptivity. Comprising a small offset of the layer thickness, the calculated absorptivity is in well agreement with the experimental results.\\

Overall, considering its negligible small thermal mass and broad spectral efficiency, we suggest that the fabricated absorber is ideal for thermal IR and THz detector applications. Furthermore, the obtained data indicate the potential use of ultrathin Au below its percolation to be used as 'antireflection' coating.


\begin{acknowledgments}
We wish to acknowledge the support of Philipp Altpeter and Philipp Paulitschke for giving access to the clean room facilities of LMU Munich. Furtherstill, we want to thank our technicians Sophia Ewert, Michael Buchholz, and Patrik Meyer for their ongoing support. Special thanks to Christoph Eisenmenger-Sittner from the physical department for condensed matter at TU Wien for providing access to their homemade four-probe station. And we would like to thank Michael Feiginov for the discussions. D. H\o j and U.L. Andersen acknowledge support from the Villum foundation and the Danish National Research Foundation. This work has received funding from the European Research Council under the European Union's Horizon 2020 research and innovation program (Grant Agreement-716087-PLASMECS).
\end{acknowledgments}

%

\appendix

\section{Optical model}\label{app:OpticalModel}
As mentioned in the text, the optical model is based on \cite{Centurioni2005}. With the assumptions mentioned in the text the multi-layer membrane can optically be described as
\begin{equation}
    \begin{bmatrix}
        E^+_{0R}
        \\
        E^-_{0R}
    \end{bmatrix}
    = \mat{S}
    \begin{bmatrix}
        E^+_{(n+1)L}
        \\
        E^-_{(n+1)L}
    \end{bmatrix}
    \ ,
\end{equation}
where $E^+_i$ and $E^-_i$ is the electric field associated with a positive and negative going direction, respectively. The subscript $i$ indicates the E-field's position in the layered system with $0R$ and $(k+1)L$ being just before the first layer and just after the last layer, respectively. $\mat{S}$ is the system matrix. For a vacuum-Au-Si$_x$N$_y$-vacuum system using the subscripts $v$, $m$, $d$, $v$, respectively, this is defined as
\begin{equation}
    \mat{S} = \mat{I}_{vm}\mat{L}_{m}\mat{I}_{md}\mat{L}_{d}\mat{I}_{dv}
    \ ,
    \label{eqn:SystemMatrix}
\end{equation}
where $I_{ij}$ and $L_{j}$ describes the wave propagation through the interface and film, respectively, and defined as
\begin{align}
    \mat{I}_{ij} &= \frac{1}{t_{ij}}
    \begin{bmatrix}
        1 & r_{ij} \\
        r_{ij} & 1
    \end{bmatrix}
    \\
    \mat{L}_{j} &=
    \begin{bmatrix}
        \exp(-\mathrm{i}\beta_j) & 0 \\
        0 & \exp(\mathrm{i}\beta_j)
    \end{bmatrix}
    \ ,
\end{align}
and
\begin{equation}
    r_{ij} = \frac{N_j - N_i}{N_j + N_i}
    \quad
    t_{ij} = \frac{2N_i}{N_j+N_i}
    \quad
    \beta_j = \frac{2\pi d_jN_j}{\lambda}
    \ ,
\end{equation}
where $N_i$ is the complex refractive index of a material layer, $d_j$ its thickness, and $\lambda$ is the optical wavelength in free space. From $\mat{S}$ it is possible to estimate the overall reflection and transmission coefficients $r = S_{21}/S_{11}$ and $t=1/S_{11}$, respectively, from which the system's reflectivity $R$ and transmittance $T$ are defined as
\begin{equation}
    R = \left|r^2\right|
    \quad , \quad
    T = \left|t^2\right|
    \ .
\end{equation}
Note that these equations are only valid with the assumptions given in the text and using vacuum as the first and final material layer. For a system only containing Si$_x$N$_y$, equation \ref{eqn:SystemMatrix} reduces to
\begin{equation}
    \mat{S} = \mat{I}_{vd}\mat{L}_{d}\mat{I}_{dv}
    \ .
\end{equation}

\end{document}